\documentclass[12pt]{iopart}

\usepackage{bm}
\usepackage{verbatim}
\usepackage{color}
\usepackage{graphicx}
\DeclareMathAlphabet{\mathpzc}{OT1}{pzc}{L}{it}

\begin{document}

\title[Waves in time-varying ${\cal PT}$-symmetric structures]
{Instantaneous modulations in time-varying complex optical potentials}

\author{Armen G Hayrapetyan$^1$, S P Klevansky$^{2,3}$ and J\"org B G\"otte$^{1,4}$}

\address{$^1$Max-Planck-Institut f\"ur Physik komplexer Systeme, N\"othnitzer Str. 38, 01187 Dresden, Germany}
\address{$^2$Institut f\"ur Theoretische Physik, Universit\"at Heidelberg, Philosophenweg 12, 69120 Heidelberg, Germany}
\address{$^3$Department of Physics, University of the Witwatersrand, Johannesburg, South Africa}
\address{$^4$School of Physics and Astronomy, University of Glasgow, Glasgow G12 8QQ, U.K.}

\ead{armen@pks.mpg.de, armen.hayr@gmail.com}
\vspace{10pt}

\begin{abstract}
We study the impact of a spatially homogeneous yet non-stationary dielectric 
permittivity on the dynamical and spectral properties of light. 
Our choice of potential is motivated by the interest in ${\cal PT}$-symmetric systems 
as an extension of quantum mechanics. 
Because we consider a homogeneous and non-stationary medium, ${\cal PT}$ symmetry reduces to time-reversal symmetry in the presence of balanced gain and loss. 
We construct the instantaneous amplitude and angular frequency of waves within the framework of Maxwell's equations and demonstrate the modulation of light \textit{amplification} and \textit{attenuation} associated with the well-defined  
temporal domains of \textit{gain} and \textit{loss}, respectively. 
Moreover, we predict the \textit{splitting} of extrema of the angular frequency modulation and demonstrate 
the associated shrinkage of the modulation period.
Our theory can be extended for investigating
similar time-dependent effects with matter and acoustic waves in ${\cal PT}$-symmetric structures.
\end{abstract}

%

\vspace{2pc}

%
%
%
%

\section{Introduction}

During the past years, a new class of Hamiltonians 
has been widely investigated, which extends quantum mechanics
from the Hermitian into the non-Hermitian (complex) domain~\cite{Bender:07}.
Despite the lack of Hermiticity, Bender \textit{et al.} have shown in their seminal papers
that a Hamiltonian can have real eigenspectra if it possesses so-called parity-time (${\cal PT}$)
symmetry~\cite{Bender:98,Bender:99}. Such a symmetry means there is invariance of the theory 
under parity (spatial) reflection ${\cal P}$: $\hat{p} \rightarrow -\hat{p}$, $\hat{x} \rightarrow - \hat{x}$,
and time reflection ${\cal T}$:  $\hat{p} \rightarrow -\hat{p}$ ($t \rightarrow - t$), $i \rightarrow - i$, $\hat{x} \rightarrow \hat{x}$,
where $\hat{p}$ and $\hat{x}$ are the momentum and position operators, respectively, 
while $t$ is the time coordinate and $i$ is the imaginary unit.
This combined ${\cal{PT}}$ symmetry leads to subtle changes in the unitary evolution of 
the system and modification of the inner product~\cite{Bender:02,Bender:04,Mosta:02a,Mosta:02b}.
As $\cal{PT}$ symmetry represents an extension of quantum mechanics, it is nowadays 
used in various different contexts, such as quantum reflection~\cite{Mosta:09,Dasarathy:13,Sinha:13} and
chaos~\cite{West:10}, and 
has even been generalized to fermionic~\cite{Jones-Smith:10,Bender:11}, gyrotropic~\cite{Lee:14,Li:16} and magnetic
systems~\cite{Lee:15}.

Although the concept of ${\cal PT}$ symmetry was originally introduced in 
quantum mechanical systems, one has found 
experimental evidence and also a wide range of applications in classical optics. 
In 2010, R\"uter \textit{et al.} were the first to realize a ${\cal PT}$-optical coupled 
system that involves well-defined regions with \textit{gain} and \textit{loss} regimes,
inherent to the \textit{complex-valued} refractive index~\cite{Rueter:10}.
Such an extension of the concept of spacetime reflection into the classical domain stems from the 
works of El-Ganainy \textit{et al.} \cite{El-Ganainy:07}
and Makris \textit{et al.} \cite{Makris:08,Makris:10},
who have employed the similarity between the Schr\"odinger and a scalar approximation of 
Maxwell's equations
to describe the dynamics of light beams in ${\cal PT}$-symmetric optical lattices.
There have been further theoretical~\cite{Berry:08,Yu:12,Sukhorukov:12,Goette:14} and 
experimental~\cite{Regensburger:12,Eichelkraut:13,Principe:15}
studies dealing with the implementation of the parity-time reversal symmetry in optics
especially relevant for the development
of new artificial structures and materials
[see also the recent review paper~\cite{Zyablovsky:14a} and references cited therein].

${\cal PT}$-symmetric structures have mainly been investigated
in the spatial domain (that is, for time-independent complex potentials)
and little attention has been paid to the study of the non-stationary 
regime. The significance of consideration of the \textit{time-dependent} potentials
both in the quantum~\cite{Faria:06,Faria:07,Moiseyev:11,Wu:12,Valle:13} and 
classical theories~\cite{Longhi:09,El-Ganainy:12} arises from the attempt
to examine the full time evolution of the system.
Despite the recent works, however, there are no rigorous analytical studies 
of wave equations 
with non-stationary complex potentials possessing time reflection symmetry. 
Given that the energy (frequency) and time are conjugate variables, 
a successful solution of ${\cal PT}$-symmetric time-dependent
Maxwell's equations would constitute a complete characterization
of the dynamical and spectral features of light.

The purpose of this paper is to study
the dynamics of light in 
time-dependent optical potentials having ${\cal PT}$ symmetry.
We consider a spatially homogenenous system, for which ${\cal PT}$ reduces to symmetry under time reversal, albeit in the presence of both gain and loss.
In view of this, we calculate both the instantaneous amplitude and angular frequency
of waves and show how the complex-valued dielectric permittivity 
controls the light in the temporal domain.
The resulting amplification and attenuation of the amplitude is demonstrated to be
associated with well-defined regimes of gain and loss, respectively.
A comparison with the angular frequency modulation
by a real permittivity is provided to reveal the impact of ${\cal PT}$-symmetric
potentials in that we observe (i) splitting of the extrema and
(ii) a shrinkage of the frequency modulation period.
Moreover, a particular emphasis is placed on studying modulations of light amplification
and attenuation for experimentally accessible values of the modulated permittivity.

The paper is organized as follows. In Section~\ref{theory}
we briefly discuss ${\cal PT}$ symmetry in optics by considering space-independent
but time-varying modulations of a complex-valued dielectric permittivity.
For such a modulated permittivity, symmetric under time-reversal, we construct an analytical 
solution to Maxwell's equations when the modulation rate is assumed to be much 
smaller than the wave frequency. This solution is further exploited in Section~\ref{dynamics}
to derive and analyze the
instantaneous amplitude and angular frequency of light. Peculiar properties, such as
modulations of amplitude amplification and attenuation as well as 
the splitting of frequency extrema and shrinkage of frequency modulation, are discussed in detail.
Finally, conclusions are given in Section~\ref{conclusions}.
\section{Solution to Maxwell's equation for space-independent but
time-varying dielectric permittivity obeying ${\cal PT}$ symmetry}
\label{theory}

\begin{figure}[htbp]
\centerline{\includegraphics[width=.6\columnwidth]{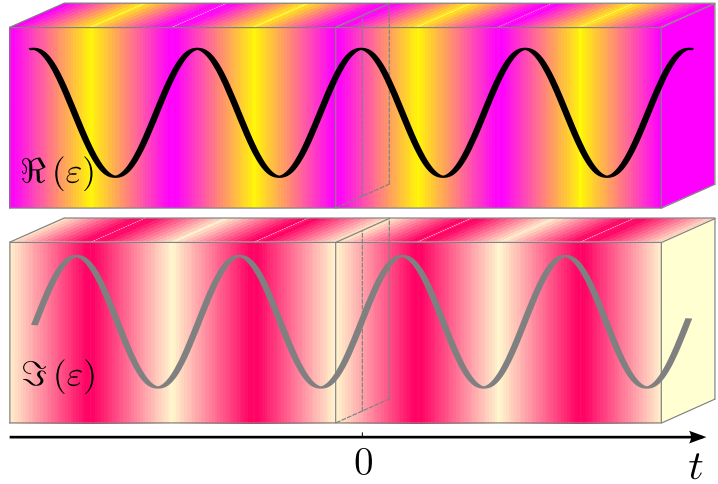}}
\caption{Symmetric real, $\Re \left( \varepsilon \right)$, and anti-symmetric imaginary, $\Im \left( \varepsilon \right)$, 
parts of the time-dependent
dielectric permittivity.}
\label{figure1}
\end{figure}

We start with a brief discussion of general properties of classical optical systems
possessing ${\cal PT}$ symmetry. 
Building on the formal equivalence of the Schr\"odinger equation with the paraxial Helmholtz (Maxwell) equation, we identify the complex refractive index $n = \Re \left( n \right) + i \Im \left( n \right)$ as the optical potential,
the real ($\Re \left( n \right)$) and imaginary ($\Im \left( n \right)$) parts of which are, 
correspondingly, even and odd functions of spacetime 
coordinates to ensure the ${\cal PT}$ invariance of the theory~\cite{Rueter:10,El-Ganainy:07,Makris:08,Makris:10}.
Likewise, since for non-magnetic structures the real ($\Re \left( \varepsilon \right)$) and
imaginary ($\Im \left( \varepsilon \right)$) parts of the dielectric permittivity are defined 
via $\Re \left( \varepsilon \right) = \left[ \Re \left( n \right) \right]^2 - \left[ \Im \left( n \right) \right]^2$
and $\Im \left( \varepsilon \right) = 2 \Re \left( n \right) \Im \left( n \right)$,
the symmetry, $\Re \left[ \varepsilon \left( x, t \right) \right] = \Re \left[ \varepsilon \left( - x, - t \right) \right]$,
and anti-symmetry, $\Im \left[ \varepsilon \left( x, t \right) \right] = - \Im \left[ \varepsilon \left( -x, -t \right) \right]$, 
relations guarantee that the full wave equation, without any approximation,
remains invariant under the parity-time transformation~\cite{Longhi:10,Golshani:14}.
Throughout this work, we place our emphasis on the \textit{temporal} domain and 
consider a spatially homogeneous yet \textit{time-dependent}
dielectric permittivity, $\varepsilon \left( t \right)$.
For the sake of illustration we choose
\begin{eqnarray}
\label{epsilon}
	\varepsilon \left( t \right) & \equiv & \varepsilon \left( \tau \right) \,\, = \,\,
	 \tilde\varepsilon + \varepsilon_1 \cos^2 \left( \tau \right)
	+ i \varepsilon_2 \sin \left( 2 \tau \right) \, ,
\end{eqnarray}
as a modulated time-dependent optical potential [c.f.~Fig.~\ref{figure1}],
similar to its spatial counterpart as discussed in Refs.~\cite{Makris:08,Makris:10}~\footnote{ 
Note that this choice of the frequency-independent permittivity,
being the analogue of the refractive index used
in~\cite{Makris:08} in position space, 
does not satisfy the standard Kramers-Kronig relations, i.e.,
the medium is not subject to the causality principle.
$\varepsilon \left( t \right)$ is valid for all time (that is, does not vanish
for negative times) so that \textit{a priori}
no restrictions on the past are necessary. }.
In Eq.~(\ref{epsilon}), $\tilde\varepsilon$ is the background dielectric constant, 
$\varepsilon_1$ represents the amplitude of the real profile of the potential, 
whereas $\varepsilon_2$ describes the strength of 
the gain/loss periodic distribution. Moreover,
$\tau \equiv b t$ is a dimensionless time, where $b \geq 0$ acts as a scaling factor
and indicates the rate (i.e., the frequency) of modulation of the permittivity,
that we assume to occur slower than the oscillations of the wave.
This is reminiscent of the similar form of modulation adopted in 
Refs.~\cite{Makris:08,Makris:10} for the spatial case
and could be experimentally realized by 
utilizing electro-optical systems~\cite{Segev}.
Note that we treat the amplitudes $\varepsilon_1$ and $\varepsilon_2$ as signed 
quantities in Eq.~(\ref{epsilon}), though in general the behaviour is not symmetric under 
an exchange of the sign of the amplitudes.

In order to investigate dynamical and spectral properties 
of light in non-stationary ${\cal PT}$-symmetric 
structures, we derive an exact second order differential equation from Maxwell's equations~\cite{Jackson:01}
for the electric displacement vector~$\bm{\mathcal{D}}$,
\begin{eqnarray}
\label{Maxwell}
	\Delta \bm{\mathcal{D}} \left( \bm r , t \right)  - 
	\frac{\varepsilon \left( t \right)}{c^2} \frac{\partial^2}{\partial t^2} \bm{\mathcal{D}} 
	\left( \bm r ,t \right) & = & 0 \, ,
\end{eqnarray}
which is valid for an arbitrary shape of the time-dependent dielectric permittivity~\cite{Hayrapetyan:16}.
Here, $\Delta$ is the Laplace operator and $c$ is the speed of light in vacuum
(for the theory of non-stationary electromagnetics see, e.g., 
Refs.~\cite{Shvartsburg:05,Nerukh:12,Sivan:16a}).
Without the modulation rate, i.e., when $b = 0$, 
the standard linear dispersion relation $\bm k  = \omega \sqrt{\varepsilon \left( 0 \right)} \hat{\bm{k}} /c$
holds, where $\varepsilon \left( 0 \right) \equiv 
\tilde\varepsilon + \varepsilon_1$ is the permittivity 
at $\tau = 0$ and $\hat{\bm{k}}$ is the unit vector in the direction of propagation.
In the presence of modulation (\ref{epsilon}),
both the amplitude and the angular frequency of light
undergo a time-dependent modification governed by Eq.~(\ref{Maxwell}). 
Accounting for this \textit{instantaneous} effect we seek a solution of 
Eq.~(\ref{Maxwell}) by making the ansatz 
\begin{eqnarray}
\label{Ansatz}
	\bm{\mathcal{D}} \left( \bm r , t \right) & = &
	\hat{\bm{u}} e^{ i \left( \bm k \cdot \bm r -  \omega t \right)} \, {\cal F} \left( \tau \right) \, ,
\end{eqnarray}
which reflects the spatial homogeneity of the permittivity.
Here, $\hat{\bm{u}}$ is the unit vector along the polarization direction, while 
the complex-valued `amplitude' ${\cal F}$ 
describes the influence of the modulated
potential on the light. 
In the absence of any modulation, 
we expect to recover the free propagation of light through a uniform medium with a
constant dielectric permittivity
so that ${\cal F} = 1$.
Note that a similar (full) wave equation for the \textit{space}-dependent electric field and 
permittivity is discussed in Ref.\ \cite{Longhi:10}
for describing the so-called ${\cal PT}$-symmetric coherent-perfect-absorber laser. 
Moreover, time reversal and time-dependent wave propagation is studied in various aspects, 
such as for time-localized perturbations combined with 
spatial periodicity~\cite{Sivan:11a,Sivan:11b,Dodin:14,Sivan:16b}
and sigmoidally changing systems with
either real or complex permittivity/refractive 
index~\cite{Hayrapetyan:16,Euser:05,Harding:07,Euser:08,Mkrtchyan:10}.

Next, we insert the ansatz (\ref{Ansatz}) in Eq. (\ref{Maxwell}) and obtain
a second order linear differential equation for~${\cal F}$,
\begin{eqnarray}
\label{equationF}
	\left( \frac{b}{\omega} \right)^2 \ddot{\cal F} \left( \tau \right) - 2 i \frac{b}{\omega} \, 
	\dot {\cal F} \left( \tau \right)
	+ \left(  \frac{  \varepsilon \left( 0 \right)}{\varepsilon \left( \tau \right)} - 1 \right) 
	{\cal F} \left( \tau \right) & = & 0 \, , \quad
\end{eqnarray}
where ``dot'' refers to the derivative with respect to the dimensionless time $\tau$.
As our interest is restricted to modulations of the complex dielectric 
permittivity profile, which are slow when compared to the oscillations of light, we can adopt $b / \omega \ll 1$ and
henceforth safely ignore the first term in Eq. (\ref{equationF}). In this approximation,
the remaining first order differential equation generally determines the 
instantaneous angular frequency as 
\begin{eqnarray}
\label{inst.freq}
	\Omega \left( \tau \right)  & = &
	\omega - b \Im \left( \frac{\dot{\cal F}}{\cal F} \right)
	 \,\, = \,\,
	\frac{\omega}{2} \left[
	1+ \frac{ {\varepsilon \left( 0 \right)}  \Re \left[ \varepsilon  \left( \tau \right) \right]}
	{\left| \varepsilon  \left( \tau \right)  \right|^2} 
	\right] \, , \quad
\end{eqnarray}
for an arbitrary form of the dielectric permittivity~\footnote{This reminds us 
of the analogous definition of the \textit{local} wave vector in the spatial domain (see, e.g., Ref.~\cite{Berry:07}).}.
The exact solution of the reduced equation
when integrated from the ``initial'' time $0$ to some instant of time $\tau$ leads to
\begin{eqnarray}
\label{solution}
	{\cal F} \left( \tau \right) & = & \exp \left\{ 
	i \frac{\omega \tau}{2 b} -
         \frac{\omega {\cal C}}{2 b} \textrm{arctanh} \big( {\cal C} {\cal B} \big)
	\right\}
\\[0.1cm]
\nonumber
	& \times &
	\exp \left\{ 
         \frac{\omega {\cal C}}{2 b} \textrm{arctanh} \Big[ {\cal C} \big( {\cal B}  - 
	i {\cal A}\tan  \tau  \big) \Big]
	\right\} \, , \, \quad
\end{eqnarray}
that explicitly exhibits the ${\cal PT}$ symmetry of the displacement, 
$\bm{\mathcal{D}}^{{\cal PT}} = \bm{\mathcal{D}}$.
In Eq.~(\ref{solution}), 
the constant parameters
${\cal A}$, ${\cal B}$, ${\cal C}$ are introduced for the sake of brevity:
 ${\cal A} \equiv \tilde\varepsilon/ ( \tilde\varepsilon + \varepsilon_1 ) > 0$
carries information about the real potential, 
whereas ${\cal B} \equiv \varepsilon_2/ ( \tilde\varepsilon + \varepsilon_1 ) $
amounts to the complex-valued permittivity, being the signature of the gain/loss mechanism.
Both ${\cal A}$ and ${\cal B}$, combined with ${\cal C} \equiv 1/\sqrt{{\cal A} + {\cal B}^2} > 0$,
demonstrate the significance of the real and imaginary parts of the permittivity 
in the instantaneous character of light.
Note that any modulation vanishes if 
${\cal B} = 0$, ${\cal A} = {\cal C} = 1 $ ($\varepsilon_1 = \varepsilon_2 = 0$)
and/or $b=0$ so that we recover the anticipated free propagation of light,
as also $\lim_{b \rightarrow 0} {\cal F} = 1$. 
The solution (\ref{solution}) holds for a class of potentials
of the form (\ref{epsilon}), which is fully determined on choosing
${\cal A}$, ${\cal B}$ and one of the constants in the potential, say $\tilde{\varepsilon}$.
In addition, to mark out the range of variation of these parameters,
we insert the (approximate) solution (\ref{solution}) into the exact Eq.~(\ref{equationF})
and estimate the error for a given modulation frequency $b$
and angular frequency $\omega$ numerically. As can be easily checked, 
the relative error (that is, the term $\left( b/\omega \right)^2 \ddot{\cal F}/{\cal F}$) 
does not exceed $0.13$ for ${\cal A} \in \left[0.7, 1.7 \right]$ 
and ${\cal B} \in \left[-0.8, 0.8 \right]$ for a ratio of $b/\omega = 0.01$~\footnote{The choice
for the ratio of modulation and wave frequencies is to estimate the upper value 
of $b/\omega$, for which our approximation is accurate. 
The effects, as proposed below, remain the same for smaller values of $b/\omega$.}.
Moreover, other analytical solutions for ${\cal PT}$-symmetric (quantum mechanical)
potentials can be found in Refs.~\cite{Sinha:04,Musslimani:08}.

\section{Instantaneous characteristics of ${\cal PT}$-modulated electromagnetic waves}
\label{dynamics}

The solution (\ref{solution}) allows us to fully characterize the 
dynamics of waves in non-stationary complex potentials. Indeed, by
decoupling the real and imaginary parts of
the time-dependent component of the electric displacement, 
$e^{ - i  \omega t } {\cal F} \left( \tau \right) 
= \left| \bm{\mathcal{D}} \left( \tau \right) \right| e^{i \Phi \left( \tau \right)}$,
we obtain direct access to the profile of the instantaneous amplitude $\left| {\bm{\mathcal{D}}  \left( \tau \right)}  \right|$
\begin{eqnarray}
\label{squared.D}
	 \left| {\bm{\mathcal{D}}  \left( \tau \right)}  \right|^2 & = & \left| {\cal F} \left( \tau \right) \right|^2
	 \,\, = \,\, \exp \left\{ 
	- \omega {\cal C} b^{-1} \textrm{arctanh} \big( {\cal C} {\cal B} \big)
	\right\}
\\[0.1cm]
\nonumber
	& \times &
	\exp \left\{ \frac{ \omega {\cal C}}{2 b}
	\textrm{arctanh} \frac{2 {\cal B}}{{\cal C} \left( 
	{\cal A} + 2 {\cal B}^2 +  {\cal A}^2 \tan^2 \tau \right)}
  \right\} 
\end{eqnarray}
and also to that of the instantaneous phase
\begin{eqnarray*}
\label{phase}
	\Phi \left( \tau \right) & = & -\frac{\omega \tau }{2 b}
	-\frac{\omega {\cal C}}{4 b} \left(
	\arctan \frac{{\cal A}{\cal C} \tan \tau}{1 +{\cal B}  {\cal C} } +
	\arctan \frac{{\cal A}{\cal C} \tan \tau}{1 - {\cal B}  {\cal C} } \right) ,
\end{eqnarray*}
the first derivative of which yields the profile (\ref{inst.freq}) of the instantaneous angular frequency, 
$\Omega = - b \dot{\Phi}$, as one would expect. 
Its explicit form expressed in terms of parameters ${\cal A}$ and ${\cal B}$ is
\begin{eqnarray}
\label{inst.freqAB}
	\Omega \left( \tau \right) & = & \frac{\omega}{2} 
	\left[ 1 + \frac{{\cal A} \sin^2 \tau + \cos^2 \tau}
	{\left( {\cal A} \sin^2 \tau + \cos^2 \tau \right)^2 + {\cal B}^2 \sin^2 \left( 2 \tau \right)}
	\right] \, .
\end{eqnarray}
Again, note that when the ${\cal PT}$ modulation is ``switched off'',
the relations $\left| {\bm{\mathcal{D}} }  \right|^2 =1$, $\Phi = - \omega t$ and $\Omega = \omega$
are obtained (see Ref.~\cite{Cohen:95} for main study
methods of signals whose frequency content changes in time).
\begin{figure}[htbp]
\centerline{\includegraphics[width=1.0\columnwidth]{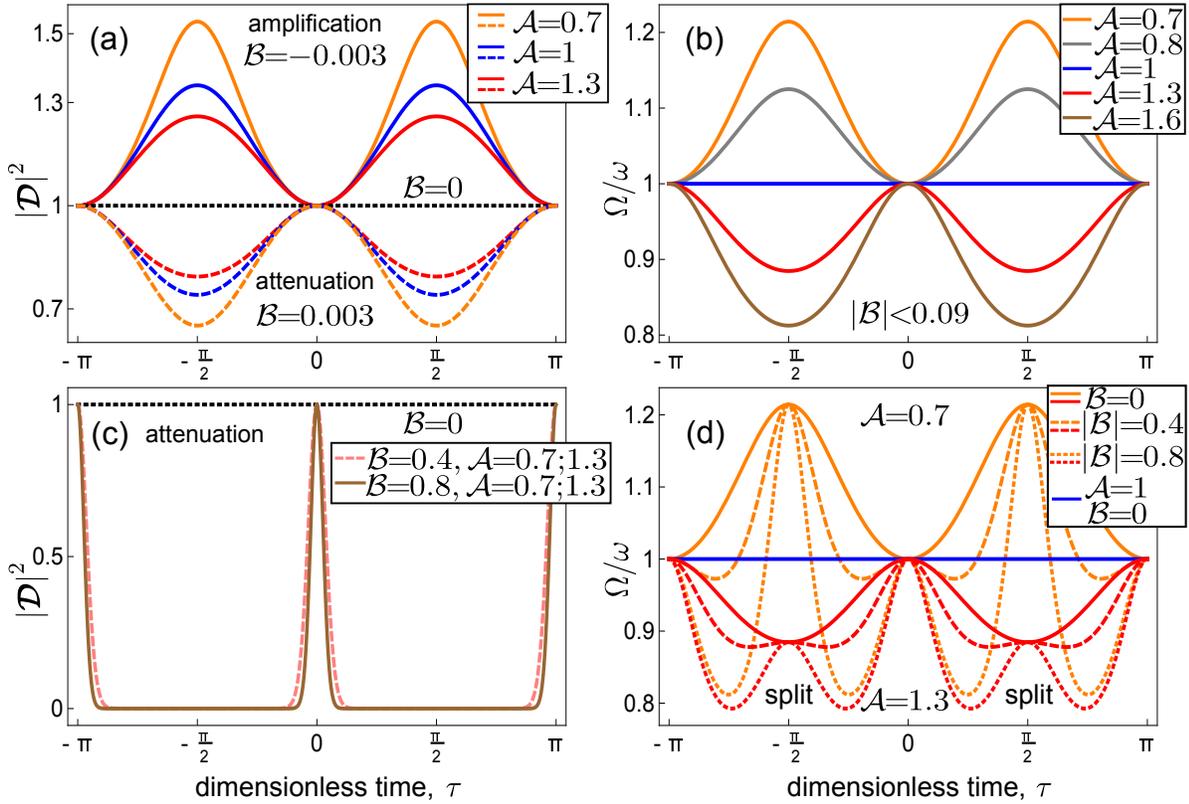}}
\caption{Instantaneous amplitude (left panel, Eq.~(\ref{squared.D})) and angular frequency (right panel, Eq.~(\ref{inst.freq}) or (\ref{inst.freqAB})) 
of light modulated by the ${\cal PT}$-symmetric dielectric permittivity $\varepsilon$, Eq.~(\ref{epsilon}), for different values
of ${\cal A}$ and ${\cal B}$. $ \left| {\cal B} \right| < \left| 1 - \cal{A} \right|/2$
corresponds to the upper panel (bellow the threshold), $ \left| {\cal B} \right| > \left| 1 - \cal{A} \right|/2$ to the lower one
(above the threshold).}
\label{figure2}
\end{figure}

For a complete description of the instantaneous amplitude and angular frequency, and in order to 
reveal their specific properties,
we determine the extrema of (\ref{squared.D}) and (\ref{inst.freqAB}) at the stationary points $t_0 \in (-\infty, \infty)$.
They are given implictly by the equations 
\begin{eqnarray}
\nonumber
	b {\cal A}^2 \left| \varepsilon \right|^2 \dot{\left| \bm{\mathcal{D}} \right|^2} & = &
	- {\cal B} \omega {\tilde{\varepsilon}}^2 \left| \bm{\mathcal{D}} \right|^2
	\sin \left( 2 \tau \right) \,\, = \,\, 0 \, ,
\\[0.1cm]
\nonumber
	2 {\cal A}^4 \left| \varepsilon \right|^4 \dot{ \Omega}  & = &
	\omega {\tilde{\varepsilon}}^2
	\sin \left( 2 \tau \right)  
	\Big\{ \left( 1 - {\cal A} \right) {\cal A}^2 \left[ \Re \left( \varepsilon \right) \right]^2 
\\
\nonumber
	& + & 4 {\tilde{\varepsilon}}^2 {\cal B}^2 \left( {\cal A} \sin^4 \tau - \cos^4 \tau  \right) \Big\} 
\\
\label{extrema.freq}
	& = & 0 \, . \quad\quad
\end{eqnarray}
From here we immediately recognize that both quantities have extrema at $\tau_0 = \pi m / 2$ provided that $m$ is
an integer whose even values, $m = 2 {\cal N}$ (with ${\cal N}$ being an integer), result in 
$\left| {\bm{\mathcal{D}}  \left( \pi {\cal N} \right)}  \right|^2 = 1$ and $\Omega \left( \pi {\cal N} \right) = \omega$.
In contrast, the odd values, $m = 2 {\cal N} + 1$, give rise to 
\begin{eqnarray}
\label{ext.squared.D}
	\left| {\bm{\mathcal{D}}  \left( \pi/2 + \pi {\cal N} \right)}  \right|^2 & = &
	\exp \left\{ 
	- \omega {\cal C} b^{-1} \textrm{arctanh} \big( {\cal C} {\cal B} \big)
	\right\} \, , 
\\[0.1cm]
\label{ext.inst.freq1}
	\Omega \left( \pi/2 + \pi {\cal N} \right) & = & \omega \left( 1 + {\cal A} \right) / \left( 2 {\cal A} \right) \, ,
\end{eqnarray}
as obtained from Eqs.~(\ref{squared.D}) and (\ref{inst.freqAB}),
respectively. Equation (\ref{ext.squared.D}) leads to the maximum possible amplitude 
amplification (${\cal B}<0$) and attenuation (${\cal B}>0$) correspondingly
linked to the well-defined temporal domains of gain and loss.
A feature of our complex and ${{\cal PT}}$-symmetric potentials is the presence of additional extremal values at
\begin{eqnarray}
\label{ext.sin.tau0p}	
	\sin^2 \tau_0^\prime & = &
	\left( 1 - 8 {\cal A} {\cal B}^2 / \sqrt{D} \right) / \left(1 - {\cal A} \right) \,
\end{eqnarray}
when the discriminant $D \equiv 16 {\cal A} {\cal B}^2 [ 4 {\cal B}^2 - ( 1 - {\cal A} )^2 ]$
of the quadratic polynomial (in $\sin^2 \tau$) appearing 
in the curly parentheses in Eq.~(\ref{extrema.freq}) is positive, that is when $2 \left| {\cal B} \right| > \left| 1 - \cal{A} \right|$.
The threshold, however, does not guarantee that the additional extrema are to be found at real values of $\tau_0^\prime$.
For this we need to specify the value of $\mathcal{B}$ further, albeit the precise criterion depends on the value of $\mathcal{A}$.
For $0 < \mathcal{A} < 1$, the criterion for additional extrema is given by $2 |{\cal B}| > \sqrt{1-\mathcal{A}}$ 
determined by the constraint $\sin^2 \tau_0^\prime > 0$,
 whereas for $\mathcal{A} > 1$, the criterion changes to $2 |{\cal B}| > \sqrt{\mathcal{A}^2 - \mathcal{A}}$
 following from the condition $\sin^2 \tau_0^\prime < 1$. 
For $\mathcal{A} = 1$, the two criteria are equivalent, and additional extrema can be expected 
from Eq.~(\ref{extrema.freq}) for
any real value of $\mathcal{B}$ at $\tau_0^\prime = (2\mathcal{N}+1) \pi/4$.

The physical reason for this difference is the sign of the modulation in the real part of the dielelectric permittivity in Eq.~(\ref{epsilon}).
For $0 < \mathcal{A} < 1$ the modulation of the real part is positive, that is, the real part varies between 
$\tilde{\varepsilon}$ and $\tilde{\varepsilon} + \varepsilon_1$, which gives rise to an increase of the instantaneous frequency as shown in Fig.~\ref{figure2}b.
For $\mathcal{A} > 1$ the signed amplitude $\varepsilon_1$ is negative and the permittivity varies between 
$\tilde{\varepsilon} - |\varepsilon_1|$ and $\tilde{\varepsilon}$, which leads to a decrease.
In both cases the position of the additional extrema is aptly described by Eq.~(\ref{ext.sin.tau0p}), which gives rise to 
a \textit{splitting} of the modulation extrema (as shown in Fig.~\ref{figure2}d)
that would otherwise be determined by Eq.~(\ref{ext.inst.freq1})
in the case when the instantaneous angular frequency is modulated by the 
conventional real permittivity.
The value of the instantaneous angular frequency at these additional extrema for a modulation by the imaginary
permittivity beyond the threshold is then given by
\begin{eqnarray}
\label{ext.inst.freq2}
	\Omega \left( \tau_0^\prime \right) & = &
	\omega \, {\cal C}^2
	\left(1 + 5 {\cal A}+ 4 {\cal B}^2
	+ \sqrt{D}/\left( 4 {\cal B}^2 \right)
	 \right)/ \, 8 \, . \quad\quad 
\end{eqnarray}
It is important to note that even though changing the time origin in Eq.~(\ref{epsilon})
does affect the formal definition of time-reversal symmetry,  
the observed effects remain the same with the only modification that the symmetry point is shifted 
for modulations of both the instantaneous amplitude, Eq.~(\ref{squared.D}), and angular frequency, Eq.~(\ref{inst.freqAB}).

Time-periodic ${\cal PT}$-symmetric optical potentials feature unusual, though expected
modulations of the instantaneous properties of the light.
Figure~\ref{figure2} illustrates the evolution of these quantities,
possessing time-reversal symmetry
($\left| {\bm{\mathcal{D}}  \left( \tau \right)}  \right|^2 = \left| {\bm{\mathcal{D}}  \left( - \tau \right)}  \right|^2$,
$\Omega \left( \tau \right) = \Omega \left( - \tau \right)$),
against the dimensionless time $\tau$ for various values of ${\cal A}$ and ${\cal B}$.
Fig.~\ref{figure2}a shows that the sign of ${\cal B}$ determines whether the overall dynamics give rise to amplification (${\cal B}<0$) or attenuation  (${\cal B}>0$), despite the fact that in both cases there are equal periods of gain and loss,
occurring, however, with different strengths.
In addition to these amplitude variations, Fig.~\ref{figure2}b shows modulations of the instantaneous angular frequency towards either higher ($0<{\cal A}<1$) or lower (${\cal A}>1$) frequencies, compared to $\omega$.
For small values of $| {\cal B} | $, the instantaneous angular frequency remains mostly unaltered (Fig.~\ref{figure2}b -- the additional maxima for $\mathcal{A} = 1$ at $\tau_0^\prime = (2\mathcal{N}+1) \pi/4$ are too small to discern on this scale) and only the amplitude experiences a modulation. 
Larger values of ${\cal B}$ lead to a vigorous modulation of the amplitude
attenuation (Fig.~\ref{figure2}c) and to
a pronounced modification of the frequency modulation profile (Fig.~\ref{figure2}d).
Unlike the ordinary modulation of angular frequency, where the modulation rate
is commensurate with the scaling factor (i.e., the modulation frequency $b$) of the real potential~\cite{Yariv:84},
in our ${\cal PT}$-symmetric structure the extrema of frequency modulation experience a \textit{split}
beyond the threshold $2 \left| {\cal B} \right| = \left| 1 - {\cal A} \right|$,
as depicted in Fig.~\ref{figure2}d. 
\begin{figure}[htbp]
\centerline{\includegraphics[width=1.0\columnwidth]{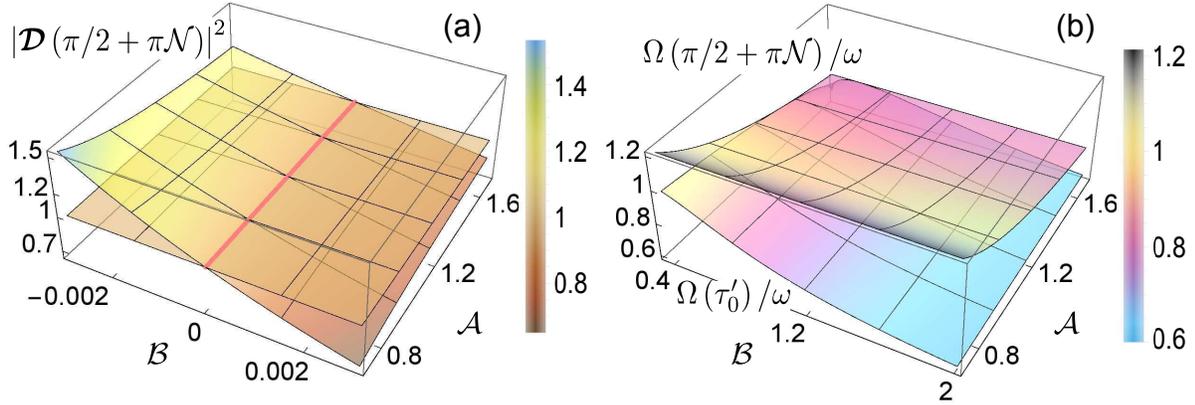}}
\caption{Extrema of the modulus square of the instantaneous amplitude ((a), Eq.~(\ref{ext.squared.D})) and angular frequency ((b), Eqs.~(\ref{ext.inst.freq1})-(\ref{ext.inst.freq2})).
The comparison is made between modulations of light via a real permittivity profile (given by the unit surface in (a) and 
the upper surface in (b)) and ${\cal {PT}}$-symmetric permittivity profiles
(given by the surface crossing unity in (a) and the lower surface in (b)).}
\label{figure3}
\end{figure}	
The global extremum turns into a local one and two new global extrema 
appear on either side such that now they occur with a \textit{shrunken}
period and the troughs of curves are shifted towards the lower frequency domain, 
quantitatively determined via Eq.~(\ref{ext.inst.freq2}).
The values of the instantaneous amplitude and angular frequency at the regular and additional extrema are 
shown in Fig.~\ref{figure3} as functions of ${\cal A}$ and ${\cal B}$,
where a comparison is made between the ${\cal PT}$-
and real-potential induced surfaces.
As the gain/loss parameter ${\cal B}$ 
changes its sign from negative to positive, the extrema of the instantaneous amplitude descend
from the region of amplification ($\left| {\bm{\mathcal{D}}  \left( \pi/2 + \pi {\cal N} \right)}  \right|^2 > 1$) to the region of
attenuation ($\left| {\bm{\mathcal{D}}  \left( \pi/2 + \pi {\cal N} \right)}  \right|^2 < 1$) for all 
values of ${\cal A}$ (Fig.~\ref{figure3}a).
By comparison, the unit surface indicates the absence
of the amplitude modulation when the imaginary part of the permittivity is ``switched off'', that is for ${\cal B} = 0$, which marks the line along which both surfaces intersect.
In contrast, as seen from Fig.~\ref{figure3}b, the extrema of the instantaneous angular frequency differ from unity
even if the imaginary part of the permittivity is zero. 
While the ${\cal B}$-\textit{in}dependent 
surface designated by $\Omega \left( \pi/2 + \pi {\cal N} \right)/\omega$ 
describes the extrema of the instantaneous angular frequency as modulated only 
by the real part of the permittivity, the surface $\Omega \left( \tau_0^\prime \right)/\omega$ always lies below 
$\Omega \left( \pi/2 + \pi {\cal N} \right)/\omega$ and
represents the split extrema due to the imaginary part of $\varepsilon$.

Until now, we have discussed how the ${\cal PT}$ modulation
of the dielectric permittivity affects the instantaneous characteristics of light
for various values of ${\cal A}$ and ${\cal B}$, as allowed for within
our approximation. Since the 
time-dependent permittivity in experimental situations can be modulated with the amplitude much less than through
the background dielectric constant~(see, e.g., Ref.~\cite{Yariv:84}), it is sensible
to discuss the impact of the permittivity in the dynamics of light 
for those values of parameters which are currently available especially in 
${\cal PT}$-coupled waveguide devices. If we consider a modulation of the dielectric permittivity 
with the real $\left| {\varepsilon_1} \right| \ll \tilde{\varepsilon}$
and imaginary $\left| {\varepsilon_2} \right| \ll \tilde{\varepsilon}$ parts and
keep terms up to first order of 
$\left| {\varepsilon_1} \right| / \tilde{\varepsilon}$ and $\left| {\varepsilon_2} \right| / \tilde{\varepsilon}$
in the parameters ${\cal A} \approx 1 - \varepsilon_1 / \tilde{\varepsilon} $,
${\cal B} \approx \varepsilon_2 / \tilde{\varepsilon}$ and 
${\cal C} \approx 1 + \varepsilon_1 / \left( 2 \tilde{\varepsilon} \right) $, we can reduce Eq.~(\ref{squared.D}) 
to an experimentally accessible form
\begin{eqnarray}
\label{squared.D.approx}
	 \left| {\bm{\mathcal{D}}  \left( \tau \right)}  \right|^2 & \approx &
	 	 \exp \left\{  \frac{ \omega}{2 b} \left[
	- 2 \textrm{arctanh} \frac{\varepsilon_2}{\tilde{\varepsilon}} 
	+ 	\textrm{arctanh} \frac{2 \varepsilon_2 \cos^2 \tau }{\tilde{\varepsilon}} \right]
	\right\} \, .
\end{eqnarray}
As seen, the instantaneous amplitude no longer depends on the value of real part of the permittivity 
modulation given by $\varepsilon_1$, but only on the 
\begin{figure}[htbp]
\centerline{\includegraphics[width=1.0\columnwidth]{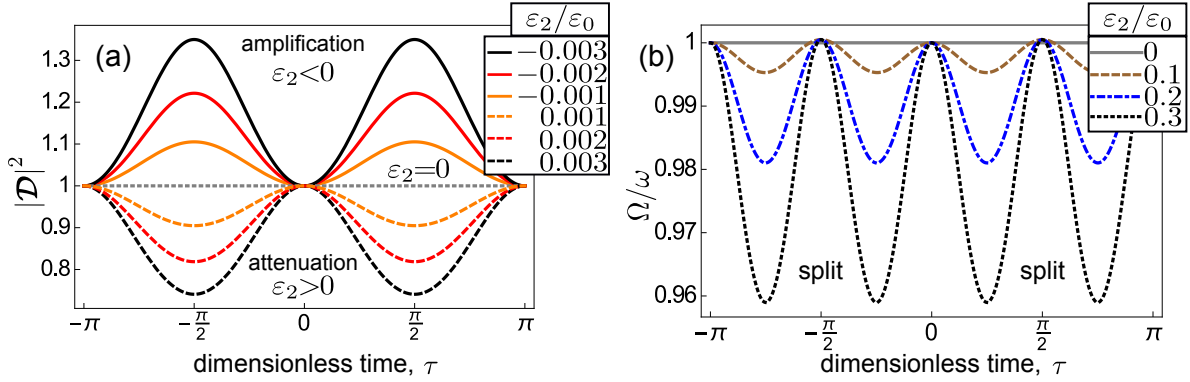}}
\caption{Instantaneous amplitude ((a), Eq.~(\ref{squared.D.approx})) and frequency ((b), Eq.~(\ref{inst.freq}))
of light modulated by the ${\cal PT}$-symmetric dielectric permittivity $\varepsilon$, Eq.~(\ref{epsilon}), for the
small value of the amplitude $\varepsilon_1$ of the real profile 
($\left| {\varepsilon_1} \right| / \tilde{\varepsilon} = 0,001$) and 
for selected values of the gain/loss strength~$\varepsilon_2$.}
\label{figure4}
\end{figure}	
gain/loss distribution strength $\varepsilon_2$, the sign of which suitably defines 
the temporal domains of gain ($\varepsilon_2 < 0$) and loss ($\varepsilon_2 > 0$),
as also shown in Fig.~\ref{figure4}a. As the magnitude of the absolute value of $\varepsilon_2$ 
increases, the amplitude modulations become more pronounced.
For a given $\left| \varepsilon_2 \right| / \tilde{\varepsilon} = 0,003$, for instance,
one can clearly distinguish between the  
increase of $\sim35$ \% and decrease of $\sim25$ \% in the amplitude modulation extrema
[the black solid and dashed curves].
Such an asymmetry in the gain and loss domains is due to the fact that Eq.~(\ref{squared.D.approx}) 
changes its form under the change of the sign of $\varepsilon_2 $.
In contrast to the amplitude modulations, that distinctly occur already for small values of $\varepsilon_2 $, 
the modulations of the IAF are driven only by the real profile since the gain/loss strength 
contributes with the second-order term in the denominator of Eq.~(\ref{inst.freq}).
However, if we allow for a strong ${\cal PT}$ coupling by 
setting $\left| \varepsilon_2 \right| / \tilde{\varepsilon} < 1$,
but keep the same restriction for the real amplitude,
$\left| \varepsilon_1 \right| / \tilde{\varepsilon} \ll1$, the instantaneous frequency will be modulated 
beyond the threshold, and therefore, the split of extrema and associated shrunken periods can
clearly be seen
at $\tau_0^\prime = (2\mathcal{N}+1) \pi/4$ since $A \approx 1$, as demonstrated in Fig.~\ref{figure4}b.
Moreover, when $\varepsilon_2 / \tilde{\varepsilon} = 0.3$, the change
in the instantaneous frequency rises by $4$ \%, as compared to the initial frequency $\omega$.

%
%

%
%
%
%
\section{Conclusions}
\label{conclusions}

In conclusion, we have examined the impact of the time-dependent
${\cal PT}$-symmetric dielectric permittivity in the dynamical and spectral features 
of light.
In our formalism, we have shown that the ${\cal PT}$ modulation of  
light is associated with the well-defined temporal domains of gain 
(${\cal B} < 0$) and loss (${\cal B} > 0$). 
We have also determined two different criteria for the splitting 
of the extrema of the angular frequency modulation to occur and we have
demonstrated the shrinkage of the modulation period.
Both the split and shrinkage are general,
inherent features of time-dependent complex potentials. 
A direct manifestation of time reflection symmetry in our particular non-stationary
structure is always evident. Such effects 
warrant more detailed future investigations with different 
time-dependent ${\cal PT}$-symmetric potentials.
It is true that while the predicted curves on Fig.~\ref{figure4} could 
be observed experimentally with techniques available nowadays, 
the curves on Fig.~\ref{figure2} are likely to only be accessed in future.

We should mention that our choice of the 
permittivity suggests that the causality principle is
not fulfilled, i.e., the parameters of the permittivity
do not follow conventional dispersion constraints, as represented by the Kramers-Kronig relations. 
Given the known inconsistencies with the
Kramers-Kronig relations for ${\cal PT}$-symmetric systems~\cite{Zyablovsky:14b,Wang:14,Baranov:15,Wang:15a}
and other artificial metamaterials with complex-valued 
permittivity~\cite{Markel:08,Silveirinha:11,Alu:11,Liu:13,Dirdal:15},
analogous relations must be constructed for frequency-independent
${\cal PT}$-symmetric time-varying dielectric permittivities,
relaxing the strict assumptions of the causality.

Although the theory developed here can be readily expanded
for studying similar ${\cal PT}$-induced effects for matter \cite{Hayrapetyan:13a,Longhi:14b,Hayrapetyan:14,Hayrapetyan:15} and 
acoustic~\cite{Mkrtchyan:10,Hayrapetyan:13b,Zhu:14,Wang:15b} waves, it also
has indirect implications for time-dependent coupling in mechanical systems~\cite{Bender:13,Bender:14}.
The consideration of the analogous theory for modified ${\cal PT}$ 
symmetries \cite{Guo:09,Ornigotti:14,Teimourpour:14} would be of great interest.

\section*{Acknowledgements}

A.G.H. acknowledges the hospitality of the Heidelberg
Graduate School of Fundamental Physics. Discussions with Benjamin Stickler,
Koen van Kruining and Robert Cameron are highly appreciated.
We would also like to acknowledge Prof. Sir Michael
Berry's comments at the conference on Correlation Optics 2015 in Ukraine. 
We acknowledge financial support from the UK Engineering and Physical Science
Research Council via the grant EP/M01326X/1 and the FET grant 295293 of the 
7th framework programme of the European Commission.

\end{document}